\documentstyle[12pt]{article}
\catcode`@=11   % WATCH OUT !!!!! This must be reset in the end !!!!!
\def\IR{\hbox{I\kern-5pt R}}

\newcommand{\eq}{\begin{equation}}
\newcommand{\en}{\end{equation}}
\newcommand{\eqn}{\begin{eqnarray}}
\newcommand{\enn}{\end{eqnarray}}

\newcommand{\beq}{\begin{equation}}
\newcommand{\eeq}{\end{equation}}

\def\CP#1{\relax\ifmmode\IP^{#1}\else\IP$^{#1}$\fi}

\def\IP{\relax\leavevmode{\rm I\kern-.18em P}}

\def\Ione{\relax\leavevmode{\rm 1\kern-3pt l}}
\def\sgn{\mathop{\operator@font sign}\nolimits}

\let\7=\widetilde

\def\ZZ{\relax\leavevmode
                         \ifmmode\mathchoice
                         {\hbox{\sf Z\kern-.4em Z}}
                         {\hbox{\sf Z\kern-.4em Z}}
                         {\lower.9pt\hbox{\scriptsize\sf Z\kern-.36em Z}}
                         {\lower1.2pt\hbox{\tiny\sf Z\kern-.36em Z}}
                          \else{\sf Z\kern-.4em Z}\fi}
\def\RR{\relax\leavevmode
                         \ifmmode\mathchoice
                         {\hbox{\sf R\kern-.4em R}}
                         {\hbox{\sf R\kern-.4em R}}
                         {\lower.9pt\hbox{\scriptsize\sf R\kern-.36em R}}
                         {\lower1.2pt\hbox{\tiny\sf R\kern-.36em R}}
                          \else{\sf R\kern-.4em R}\fi}

\def\resetby#1#2{\@addtoreset{#2}{#1}}
\def\seceq{\@addtoreset{equation}{section}% Numbers Eq.s within Sect.s
                   \def\theequation{\thesection.\arabic{equation}}} % (Sect.Eq)

\def\Label#1{\label{#1}%
               \smash{\hbox to0pt{\raise1ex\hbox{\tiny[#1]}\hss}}}
\def\noLabels{\let\Label=\label}

\thicklines     

\reversemarginpar
\catcode`@=12                   % You see ?
\begin{document}

\begin{titlepage}
\begin{flushright}
VPI-IPPAP-02-03\\
hep-th/0202173\\
\end{flushright}

\begin{center}

{\large\bf{Nambu Quantum Mechanics: A Nonlinear Generalization of
Geometric Quantum
Mechanics}}\\[10mm]
{\bf Djordje Minic\footnote{e-mail: dminic@vt.edu} and Chia-Hsiung
Tze\footnote{e-mail: kahong@vt.edu} } \\[5mm]
                    {\it Institute for Particle Physics and Astrophysics}\\
                  {\it Department of Physics}\\
                  {\it Virginia Tech}\\
                  {\it Blacksburg, VA 24061}\\[10mm]

{\bf ABSTRACT}\\[3mm]
\parbox{4.8in}{We propose a generalization of the standard geometric
formulation of
quantum mechanics, based on the
classical Nambu dynamics of free Euler tops. This extended quantum mechanics
has in lieu of the standard exponential time evolution, a nonlinear
temporal evolution given by Jacobi elliptic functions. In the limit
where latter's
moduli parameters are set to zero, the usual
geometric formulation of quantum mechanics, based on the Kahler structure
of a complex projective Hilbert space,
is recovered. We point out various novel features of this
extended quantum mechanics, including its geometric aspects. Our approach
sheds a new light on the problem of quantization of Nambu dynamics.
Finally, we argue that the structure of this nonlinear quantum mechanics
is natural from the point of view of string theory.}
\end{center}
\end{titlepage}

Generalizations of quantum mechanics are
difficult. One idea which has often been exploited is to make the
Schr\"{o}dinger equation non-linear (see for example \cite{weinberg}).
Another approach is to extend the quantum phase space from the usual
complex projective
space to an arbitrary Kahler manifold. Yet another avenue
is to enlarge
complex quantum mechanics by changing the coefficients on
the Hilbert
space to quaternions
\cite{adler} or octonions
\cite{gursey, okubo}.  In this article a more radical approach to this 
question is
investigated. What is proposed is a modification of the kinematics
and dynamics, the
very symplectic and Riemannian structure of geometric complex quantum
mechanics.

Any search for generalizations of quantum mechanics has
to have a well defined motivation.
One possible general starting point is provided by the observation that the 
evolution of fundamental physical theories, characterized by
appearance of
new dimensionful parameters (new constants of nature), can be 
mathematically understood from the point of view of deformation theory 
\cite{faddeev}. In particular,
relativity theory, quantum
mechanics and quantum field theory can be
understood mathematically as
deformations of unstable structures \cite{flato}\footnote{An algebraic 
structure is termed stable (or
rigid) for a class of
deformations if any deformation in this class leads to an equivalent
(isomorphic)
structure.}.
An example of an unstable
algebraic structure is non-relativistic classical mechanics.
By deforming an unstable structure, such as classical non-relativistic 
mechanics,
via dimensionful deformation parameters, the speed
of light $c$ and the
Planck constant $\hbar$, one obtains new stable structures - special 
relativity and
quantum mechanics.
Likewise, relativistic quantum mechanics (quantum field theory)
can be obtained through
a double ($c$ and $\hbar$) deformation. It is natural to expect that
there is a further deformation via one more dimensionful constant, the 
Planck length
$l_P$. The resulting structure could be expected to form a stable 
structural basis for a quantum theory of gravity.

A closely related idea has appeared in open string field theory, as 
originally formulated by Witten \cite{witten}. There, the
deformation parameters are $\alpha'$ and $\hbar$. The classical
open string field theory lagrangian is based on
the use of the string field (which involves an expansion to all orders in
$\alpha'$) and a star product which is defined in terms of the
world-sheet path integral,
also involving $\alpha'$. The full quantum string field theory is thus, in 
principle,
an example of a one-parameter ($\alpha'$) deformation of quantum 
mechanics\footnote{Similarly, one can also intuit that
string theory calls for a generalization of quantum
mechanics from the existence of the minimal length uncertainty relations in 
the framework of perturbative string theory
\cite{tatsu}.}.

In this letter we lay the basis for a generalized quantum
mechanics based on the
classical Nambu dynamics \cite{nambu} of Euler's asymmetric top. This
    Nambu quantum
mechanics, naturally
possesses besides Planck constant, new deformation parameters.  One of its
defining experimental signatures is a nonlinear time
evolution generated by Jacobian elliptic functions, as compared to the standard
exponential time evolution of standard quantum mechanics. The new deformation
parameters are given by the moduli of the elliptic functions.
In the limit when these are set to zero, the usual geometric
formulation of quantum
mechanics, based on the Kahler structure of the space of rays in a
complex Hilbert
space, is recovered.
We point out various features of and issues raised by this
extended quantum mechanics, including its nonstandard geometric aspects. Our
approach sheds a new light on the problem of quantization of Nambu
dynamics and is
natural from the point of view of string theory.

We begin with a brief recapitulation of the geometric formulation of
quantum mechanics as originally formulated by Kibble \cite{kibble} (for
reviews of this approach consult \cite{heslot, ashtekar,
anandan, hughston}). This geometric setting will be a springboard for our
attempt to go behond standard quantum mechanics.  The basic observation is that
pure states of a quantum mechanical system correspond to rays in a
complex linear
Hilbert space
${\cal{H}}$. The latter can also be seen as a real vector space with a
complex structure $J$. So the Hermitian inner product of two states $<\psi|$
and $|\phi>$ in ${\cal{H}}$ can be split into its real and imaginary parts :
\eq
{<\psi|\phi> = g(\psi,\phi)  + i \omega(\psi,\phi) =
\delta_{ij}\psi_i\phi_j + i \epsilon_{ij}\psi_i\phi_j}.
\en
with $i, j$=1,2, labelling the real and imaginary components of
$ \psi$ and $\phi$ . The metric $g$ is the scalar product and the
antisymmetric
$\omega$ is a symplectic 2-form, they are related through $J$ as
$g(\psi,\phi) = \omega(\psi,J\phi)$. The triple $(g,\omega, J)$
makes ${\cal{H}}$ a Kahler space. Thus the curved space of rays of
${\cal{H}}$, called the projective Hilbert space ${\cal{P}}$,
is the $quantum$ phase space and has the complex geometry of a Kahler
manifold. Pure states of the quantum system are represented as
points of the manifold
${\cal{P}}$, which is endowed
with natural symplectic and Riemannian structures. This symplectic
structure encodes
the symplectic structure that survives in the classical limit. One
notes that the Riemannian structure, with which the complex Kahler
structure is said to be compatible, is
absent in the classical phase space and is in fact a key
ingredient of geometric quantum theory as it encodes the  information about
pure quantum mechanical properties, such as the measurement process and
Heisenberg's uncertainty relations.
Up to numerical factors, Planck constant is given by the inverse of the
constant holomorphic sectional curvature of ${\cal{P}}$. Observables
${A = <\hat{A}>}$,
defined as the expectation value of a hermitian linear operator $\hat{A}$,
correspond to real valued
differentiable functions on ${\cal{P}}$. The derivative of such a
Kahlerian function A vanishes
at an ``eigenstate'' with the value of A at such a point being the
corresponding ``eigenvalue''.
The evolution of states (the Schr\"{o}dinger  equation) represents a
symplectic flow generated
by a Hamiltonian.

More explicitly,
let us consider a pure state $ \psi = \sum_a e_a \psi_a$, where the
$\psi_a$ are the generalized Fourier components of $ \psi $ in an
orthonormal eigenbasis
$\{e_a\}$ of the Hamiltonian $H$ of a given system. Also, setting for
convenience Planck
constant equal to 1, we let
$q^a =
\sqrt{2} Re \psi_a$ and
$p_a = \sqrt{2} Im \psi_a$ with the $(q^a + ip_a)$ providing the
homegeneous coordinates
for  ${\cal{P}}$. The natural symplectic structure on ${\cal{P}}$ is then
given by the closed, nondegenerate 2-form
$\omega^{(2)} = d p_a \wedge dq^a$, $d\omega^{(2)} = 0$.
The Poisson bracket is defined as usual:
$
\{f, g\} = {\partial f \over \partial p_a }
{\partial g \over \partial q^a} -
{\partial f \over \partial q^a}
{\partial g \over \partial p_a} \equiv \omega^{AB}  {\partial f \over
\partial x^A}
{\partial g \over \partial x^B},
$
where $ \omega^{AB}$  is the inverse of the symplectic 2-form and the $x^A
= (p_a, q^a)$ form a set of canonical coordinates.  As to the
Schr\"{o}dinger equation, with ${h = <\hat{H}>}$, it now takes the
form of Hamilton's equations:
\eq
{d p_a \over dt} = \{h, p_a \}, \quad {d q^a \over dt} = \{h, q^a\}.
\en
Here $h = {1 \over 2} \sum_a [ (p^a)^2 + (q_a)^2 ] \omega_i$,
and $\omega_i$ denote eigenvalues of the given Hamiltonian $H$.
Thus for a general quantum system in the specific basis $\{e_a\}$  , the
Hamiltonian $h$
describes an infinite set of abstract free harmonic
oscillators! Thus we have
the alternative view of quantum mechanics as a
rather familiar classical Hamiltonian system, albeit one on a
(generally  infinite-dimensional) nonlinear, projective Hilbert
space
\cite {Bloch}. The Schr\"{o}dinger equation follows from the
variation of the action
$S = \int (p_a dq^a - h dt)$.
The time evolution is, of course, exponential:
$(q_a +i p_a)(t) = \exp(-i \omega_a t) (q_a +i p_a)(0)$.
In general, an arbitrary observable $O$ will evolve according to
\eq
{ d O \over dt}= \{h, O \}.
\en
It also turns out that the expectation values of commutators of
operators acting on
the Hilbert space
${\cal{H}}$ are the Poisson brackets of the corresponding Kahlerian 
functions in the geometric
formulation!

The normalization of the wave function $\psi^{*} \psi =1$, known as
the Born rule
(imposed by introducing a Lagrange multiplier into the above action)
becomes,
${1 \over 2} \sum_a [ (p^a)^2 + (q_a)^2 ] =1 $. Moreover the
points $\psi$ and
$e^{i\alpha} \psi$
are to be identified. Thus, if we take our Hilbert
space to be one of
finite (e.g. for a spin system) complex dimension $n+1$, namely
${\cal{H}} = C^{n+1}$, the  above kinematic
constraint says that the quantum phase space of rays in ${\cal{H}}$
is the complex
projective $CP(n)$. The latter is thus the base space of the complex
Hopf bundle
${S^{2n+1}}$ over $CP(n)$ (which can be realized as the coset space of
$U(n+1)/
U(1) \times U(n)$) with a U(1) fiber, the group of complex phases in
quantum mechanics.
The Riemannian metric on ${\cal{P}}$ is given by the natural Kahler metric. For
${\cal{P}}$ =
$CP(n)$, it is the well-known Fubini-Study metric, $ds_{12}^2 = (1 -
|<\psi_1|\psi_2>|^2)$.
For example, the Heisenberg uncertainty relations arise from such a
metric of the state
manifold whose local properties also lead to a generalized
energy-time uncertainty relation
\cite{anandan}.

Finally, given a curve $\Gamma$ in the projective Hilbert space
${\cal{P}}$, the geometric (Berry) phase \cite{gphase}
is given by \cite{anandan}
\eq
\int _{\Sigma} d p_a \wedge dq^a,
\en
where $\Sigma$ has as its boundary $\Gamma$. Its expression as a
symplectic area enclosed by $\Sigma$ shows that it results
solely from the geometry of the Hermitian inner product and is
independent of the Hamiltonian and the equation of motion provided
that the latter is first order in time.

In the foregoing it is clear from the perspective of geometric
quantum mechanics
that the process of quantization
is superfluous.  One takes quantum theory as given.
Quantum theory is in fact mathematically equivalent to a special type
of classical
Hamiltonian phase-space dynamics, albeit with a key difference;
namely the underlying
phase-space is not the finite dimensional symplectic phase space of
classical mechanics, but
rather the (infinite dimensional) quantum mechanical Kahler state
space itself, naturally
endowed with a compatible Riemannian metric. This formulation
   points to specific ways
of extending quantum mechanics. This is achieved  by tempering with
either the kinematics, or
the dynamics and quantum phase space of standard geometric quantum
mechanics or indeed with all
of these elements simultaneously. Next we choose to substitute the
dynamics of the above systems of non interacting harmonic
oscillators by a system of free Nambu tops and explore the
ensuing implications.

We begin then by summing up the corresponding geometrical
characteristics of the classical Nambu mechanics \cite{nambu}.
In this letter we make use specifically
of a Nambu system of order 3, namely a Hamiltonian system defined
with respect to a ternary
bracket.
Thus the fundamental analog of the
symplectic 2-form of usual Hamiltonian
dynamics is a closed non-degenerate 3-form
\cite{takhtajan}
\eq
\omega^{(3)} = d M_1 \wedge dM_2 \wedge dM_3,
\en
and the action given as an integral of the corresponding
Poincare-Cartan 2-form
\eq
S = \int M_1 dM_2 \wedge dM_3 - H_1 dH_2 \wedge dt.
\en This form of the action shows that initial and final states in
this type of Nambu dynamics
are described by loops  rather then
points, because
the integrand of the action is a two form, rather a one form, as in the
usual Hamiltonian dynamics. The fact that loops appear in the phase space of
triple Nambu mechanics thus points out to a possible application to string
theory.

The equations of motion - Nambu equations - follow from $\delta S=0$:
\eq
{d F \over dt}= \{H_1, H_2, F \}
\en
where $F$ is a function of $M_1, M_2, M_3$
and the Nambu-Poisson bracket $\{F,G,H \}$ is defined as
\eq
\{F,G,H \} = \epsilon^{ijk} \partial_{M_i} F \partial_{M_j} H 
\partial_{M_k} G .
\en
The Nambu bracket generates volume preserving diffeomorphisms on the
phase space of
the Nambu dynamics and the Liouville theorem is obeyed.
The Poisson bracket is evidently a natural ``contraction'' of the
Nambu bracket.
The latter
satisfies the following three
key conditions: skew-symmetry, the Leibniz rule and the
Fundamental Identity \cite{takhtajan, filip}, the counterpart of the Jacobi
identity obeyed
by Poisson brackets:
\begin{enumerate}
\item Skew-symmetry
\eq
\{A_1, A_2,A_3\}=(-1)^{\epsilon(p)} \{ A_{p(1)}, A_{p(2)},A_{p(3)} \},
\label{skewsymmetry}
\en
where $p(i)$ is the permutation of indices and
$\epsilon(p)$ is the parity of the permutation,
\item Derivation (the Leibniz rule)
\eq
\{A_1A_2, A_3, A_4\} =A_1\{A_2, A_3,A_4\} + \{A_1, A_3,A_4\}A_2 ,
\en
\item Fundamental Identity
\eqn
\{\{A_1, A_2, A_3\},A_4, A_5 \} +\{A_3, \{A_1, A_2,A_4\},A_5\}
\nonumber \\
+\{A_3, A_4, \{A_1, A_2, A_5\}\} =
\{A_1,A_2,\{A_3, A_4, A_5\}\}.
\enn
\end{enumerate}
We should mention that there exists an algebraic n-ary
generalization of the Nambu
bracket associated with a Nambu-Poisson manifold
\cite{takhtajan}. However in such a
scheme, the associated
fundamental identity is too restrictive in that the Nambu dynamics
on  a $n$-dimensional
manifold is then determined by $(n-1)$ conserved Hamiltonian functions.
This too large a number of
integrals of motion is clearly most unsuitable for our specific
generalization of geometric
quantum dynamics where $n$ is finite only for, say, spin systems but is
generically infinite.

Now, we observe that,
just as the simple harmonic
oscillator is the prototype classical and quantum system of the standard
Hamiltonian mechanics, the Euler
asymmetric top is the prototypical representative of Nambu's ternary
mechanics.  The time evolution of the
Euler top in Nambu mechanics is
described by two Hamiltonian functions given by the total
energy and a Casimir
invariant, the square of the angular momentum.  These two conserved
quantities are:
$
H_1 = {1 \over 2}\sum_{i=1}^{3}{1 \over I_i} (M_i)^2$ and $H_2 ={1 \over 2}
\sum_{i=1}^{3} (M_i)^2$, where $I_i$ denote the principal moments
of inertia.
It is easy to see that the Nambu equations of motion give the equations
written in ${\IR}^3$ for the asymmetric
Euler top, a rigid body fixed in the center of mass:
\eq
{d M_1 \over dt} = ({1 \over I_3}-{1 \over I_2})M_2 M_3, \quad
{d M_2 \over dt} = ({1 \over I_1}-{1 \over I_3})M_3 M_1, \quad
{d M_3 \over dt} = ({1 \over I_2}-{1 \over I_1})M_1 M_2,
\en
where we take $I_3> I_2 >I_1$ . These
equations have been reincarnated during recent decades in the celebrated
Nahm equations \cite{nahm} for the $SU(2)$ self-dual Yang-Mills field of 
relevance to theories of
extended objects such as monopoles and membranes. Specifically the Euler 
equations
for the asymmetric top
naturally describe geodesic flows on a triaxial ellipsoid
and can be solved in terms of Jacobi elliptic function \cite{landau}. For
later comparison, we elaborate briefly on this last point. The equations
for the asymmetric top
are closely linked to the SO(3) algebra, the group SO(3) being that of
proper orthogonal transformations in ${\IR}^3$.  ${\IR}^3$ can then be 
identified
with the  SO(3) Lie algebra since the latter is isomorphic to that of vectors
in
${\IR}^3$. It was shown
\cite{arnold} that the left invariant metric tensor on SO(3), compatible
with the top motion is given by
$g(X,Y)=- {1\over 2} K({\mathbf{I}}X,Y)$ where $K$ is the appropriate Killing
form, $\mathbf{I}$ is the moment of inertia operator and $X$ and $Y$ are 
tangent
vectors to SO(3). Then, in an SO(3) eigenbasis of $\mathbf{I}$ the above top
equations are the geodesics of the given left-invariant Riemannian metric of
the group SO(3), on a reduced phase space to be identified as a 2-sphere. Thus
in terms of the components of angular velocity
$\mathbf{\Omega}$, Eq(12) also reads
\eq
{d \Omega^i \over dt} + \Gamma^{i}_{jk} \Omega^j \Omega^k =0,
\en
where the Christoffel symbols are
$\Gamma^{i}_{jk} = \frac{1}{2} \epsilon_{jki} (1 - \frac{I_j - I_k}{I_i})$.
Indeed by being so embedded in SO(3) the
Euler top equations of motion become a Hamiltonian flow on the (co-)adjoint 
orbits of the action of SO(3) which are 2-spheres, $S^2$. As a
Hamiltonian system on ${\IR}^3$, the time evolution of any observable $G$ 
is given
through a Lie-Poisson bracket $\{ G, H_1\}$ :
\eq
{dG \over dt} = \{G, H_1 \} = \sum \epsilon_{ijk} M_i {\partial H_1 \over 
\partial M_j}
{ \partial G \over \partial M_k}.
\en
In fact its very form readily motivates the definition of the
triple Nambu bracket if one has another conserved quantity quadratic
in the $M_i$'s as is the case of $H_2$ for the top.  However, due to the
existence of another integral of motion,
$H_2$, which is an invariant level surface, a 2-sphere, this bracket is
degenerate. This 2-sphere is then the reduced phase space on which the
Lie-Poisson bracket is restricted and then nondegenerate.
The top equations can be written in the canonical Hamiltonian form, using
Darboux theorem \cite{kozlov}
by introducing two canonical variables $\alpha$ and $\beta$ such that
\eq
M_1 = \sqrt{H_2 -\beta^2} \sin{\alpha}, \quad
M_2 = \sqrt{H_2 -\beta^2} \cos{\alpha}, \quad
M_3 = \beta .
\en
The top equations have then the canonical form
\eq
{d \alpha \over dt} = {\partial H \over \partial \beta }, \quad
{d \beta \over dt} = - {\partial H \over \partial \alpha },
\en
where the Hamiltonian $H$ reads as follows
\eq
H = {1 \over 2} (\frac{{\sin{\alpha}}^2}{I_1} + \frac{{\cos{\alpha}}^2}{I_2})
(H_2 - \beta^2) + \frac{{\beta}^2}{I_3}.
\en
The symplectic structure is given by $d\alpha \wedge d\beta$.

Returning to the
dynamics of the Euler top, we underscore the case of the symmetric (or
Lagrange) top ($I_1=I_2$) for which
$M_3 =const$. In that case one obtains effectively a
simple harmonic oscillator with the characteristic frequency
$({1 \over I_1}-{1 \over I_3})M_3$. Also in this limit the Jacobi elliptic 
functions
become ordinary trigonometric functions.
We mention in passing that there exists an obvious generalization of
this system in the case of an n-ary Nambu bracket ($n>3$), generated
by $(n-1)$ quadratic
Nambu Hamiltonians. The resulting generalization of the Euler top can
be solved in terms of automorphic functions on a
Riemann surface of genus $(n-1)$, given the results of \cite{ratiu}.

Now we are set to state our proposed generalization of the geometric
formulation
of quantum mechanics. We call it for short - Nambu quantum mechanics.

The key to our scheme is the embedding of the standard quantum mechanics
represented  as a Hamiltonian system describing a collection of (N+1)
abstract harmonic
oscillators into a Nambu Hamiltonian system describing an abstract
collection of
(N+1) Euler tops. The exponential evolution from standard quantum
mechanics is then
immediately generalized into a time evolution described by the Jacobi elliptic
functions and the geometric formulation of quantum mechanics is naturally
generalized into a larger geometric structure, possessing extra deformation
parameters, one per top used - the modulus of the Jacobi elliptic
functions which
is the measure of the asymmetry of a top.

We consider a phase space described by real functions of
$(3N+3)$ variables or (N+1) Nambu triplets, $m_1^a, m_2^a, m_3^a$ (a
= 1,2,..N+1).
The fundamental equations of motion for a general observable
$O(m_1^a, m_2^a, m_3^a)$ are assumed to be of the Nambu type
\eq
{d O \over dt}= \{h_1, h_2, O \},
\en
where the Nambu bracket $\{F,G,H \}$ is defined as above
$
\{F,G,H \} = \epsilon^{ijk} \partial_{m_i^a} F \partial_{m_j^a} H
\partial_{m_k^a} G
$.
The summation over $a$ is understood. The generalized Schr\"{o}dinger
equation (the
Nambu-Schr\"{o}dinger equation) is given by the Nambu-Hamilton equation
${d m_i^a \over dt}= \{h_1, h_2, m_i^a\}$,
where the two Hamiltonians are
$
h_1 = {1 \over 2}\sum_{a} \sum_{i=1}^{3}{\alpha_i^a} (m_i^a)^2$
and $\quad h_2 ={1 \over 2}
\sum_{a} \sum_{i=1}^{3} (m_i^a)^2$. These equations follow from the
variation of $S = \int m^{a}_1 dm^{a}_2 \wedge dm^{a}_3 - h_1 dh_2 \wedge dt$.
Namely, the Nambu-Schr\"{o}dinger equations describe a collection of
(N+1) (which could be infinite) free abstract Euler tops :
\eq
{d m_1^a \over dt} = (\alpha_3^a-\alpha_2^a) m_2^a m_3^a,\quad
{d m_2^a \over dt} = (\alpha_1^a-\alpha_3^a) m_3^a m_1^a,\quad
{d m_3^a \over dt} = (\alpha_2^a-\alpha_1^a) m_1^a m_2^a .
\en
As is well known, these equations can be
integrated in closed form in terms of Jacobi elliptic functions
$sn, cn, dn$:
\eq
m_1^a(t) = K_1^a sn (c(t-t_0)), \quad
m_2^a(t) = K_2^a cn (c(t-t_0)), \quad
m_3^a(t) = K_3^a dn (c(t-t_0)) ,
\en
where the modulus $k$ of the Jacobi elliptic functions is given by
\eq
k^2 = \frac{\alpha_1 - \alpha_2}{\alpha_2 - \alpha_3} \frac{2h_1 - \alpha_3 
h_2}{h_2 \alpha_1 - 2h_1} .
\en
Note that in the Lagrange or symmetric top
limit when for example $\alpha_1^a=\alpha_2^a$ the Jacobi elliptic
function reduce
to the trigonometric functions
$ sn \rightarrow \sin$, $cn \rightarrow \cos$
and $dn \rightarrow 1$. To be more precise we
display the simple asymptotics of the Jacobi functions \cite{Abramovitz}:
\eqn
sn{u} &=& \sin{u} - \frac{1}{4} k^2 (u - \sin{u} \cos{u}) \cos{u} + O(k^4) \\
cn{u} &=& \cos{u} + \frac{1}{4} k^2 (u - \sin{u} \cos{u}) \sin{u} + O(k^4) \\
dn{u} &=& 1 - \frac{1}{4} k^2 \sin^{2}{u} + O(k^4) ,
\enn
where in our case $u =c(t-t_0)$.
In the  $\alpha_1^a=\alpha_2^a$ limit the Euler top
equations describe a (generally infinite)  collection of harmonic
oscillators and the time evolution is exponential.
Also $m_3^a=const$ and $m_1^a$ and $m_2^a$ can be identified with
the real and imaginary parts of the usual wave function, as described above
($m_1^a = q^a$, $m_2^a=p^a$).
This motivates the general expression for what we call the Nambu wave function
\eq
\Psi^a = \sum_i m_i^a e_i
\en
where $e_i$ are the usual
quaternions imaginary units such that $e_i e_j = - \delta_{ij} + 
\epsilon_{ijk} e_k$.
The quaternion conjugate Nambu wave function is
$\sum_i m_i^a {\bar{e}}_i = - {\bar{\Psi}}^a$. The inner
product reads
\eq
\bar{\Psi} \Phi = \delta_{ij} \Psi_{i} \Phi_{j} - \epsilon_{ijk}e_k \Psi_i 
\Phi_j
= \vec{\Psi} \cdot \vec{\Phi} - \vec{e}\cdot (\vec{\Psi} \times \vec{\Phi})
\en
where in complete correspondence with Eq.(1), the quaternionic real
part is the scalar vector product and the imaginary part is the
antisymmetric vector product. The second term in the above equation is the
quaternionic counterpart of the symplectic 2-form. It is at the
basis of the 3-form, characteristic of Nambu's original
mechanics \cite{nambu}.
   Due to the non-linear nature of  the
Nambu-Schr\"{o}dinger equation  the superposition principle
apparently no longer holds for $\Psi$ (See however ref.\cite{cirelli}).

It is also useful to introduce the following two $3 \times 3$ matrices
$L^{a}_{rs} = - L^{a}_{sr}$ ($L^{a}_{rr}=0$, $L^{a}_{12}=m^{a}_{3}$,
$L^{a}_{13}=-m^{a}_{2}$, $L^{a}_{23}=m^{a}_{1}$), and
$J^{a}_{rs}$ ($J^{a}_{rr} = \alpha^{a}_{r}$).
Then the Nambu-Schr\"{o}dinger equation can be written in the
following Lax form
\eq
{dL \over dt} = [L, (JL +LJ)].
\en
These equations are integrable (in complete analogy with the
Euler top equations). From the corresponding
Lax equations, one can deduce an infinite number of conservation
laws \cite{solitons}.

Notice that the analogs of the commutators of the Nambu quantum mechanics are
precisely given in terms of the classical Nambu bracket! Thus this 
formulation circumvents some of the well known problems encountered in the 
quantization of
the Nambu dynamics \cite{quantnambu}.

Given this abstract structure of the Nambu-Schr\"{o}dinger quantum mechanics,
it is natural to write down an appropriate operator version.
The operator form of the Nambu-Schr\"{o}dinger equations we suggest reads 
as follows
\eq
{d m_1^a \over dt} = {\cal{H}} m_2^a m_3^a - {\cal{H}} m_3^a m_2^a,
\en
\eq
{d m_2^a \over dt} = {\cal{H}} m_3^a m_1^a - {\cal{H}} m_1^a m_3^a,
\en
\eq
{d m_3^a \over dt} = {\cal{H}} m_1^a m_2^a - {\cal{H}} m_2^a m_1^a ,
\en
where the Hamiltonian is denoted by ${\cal{H}}$.
Similarly, for an operator $W$ the eigenvalue problem can defined by the 
equations
\eq
W m_2^a m_3^a - W m_3^a m_2^a = w m_1^a,
\en
\eq
W m_3^a m_1^a - W m_1^a m_3^a = w m_2^a,
\en
\eq
W m_1^a m_2^a - W m_2^a m_1^a = w m_3^a .
\en
Equations of this type appear in the theory of multidimensional determinants
\cite{gelfand}.

It is interesting to note that the Nambu-Heisenberg commutator
\eq
[P,Q,R,] \equiv PQR - PRQ + QRP - QPR + RPQ - RQP = i \hbar_{N}
\en
can have both finite and infinite dimension Hilbert space
realizations \cite{nambu, takhtajan, vt}.
This Nambu-Heisenberg relation suggests the following
cubic form of the Nambu-Heisenberg uncertainty principle $\Delta P \Delta Q 
\Delta R \sim \hbar_N$,
which is similar to the suggested generalization of the space-time
uncertainty relation in M-theory \cite{stu}.
Notice though, that it is not clear that a Hilbert space formulation of
the Nambu quantum mechanics is physically appropriate, mainly because of 
the lack of the superposition
principle.
Perhaps one should expect an appearance of the 2-Hilbert space structure
\cite{baez}\footnote{It has been argued in \cite{baez}
as well in \cite{hilbert} that background independent
formulation of quantum gravity might call for an algebraic structure larger
than the category of Hilbert spaces, in which Hilbert spaces of different
dimensionality naturally appear\cite{hilbert}.}.

Now we discuss the basics of the geometry of the Nambu quantum
mechanics. While Nambu
mechanics can be locally embedded as a constrained system in the
canonical Hamiltonian phase
space framework, we shall take it as a stand alone new mechanics.
Here tailored to our very
purpose is the formulation of \cite{pandit} for a system made up
with n Nambu triplets. It
closely parallels the characteristics of Hamiltonian systems though
there are some key
differences. Referring to \cite{pandit} for the details, it is worth gathering
the key features of that
formulation. The counterpart of standard symplectic phase space is the
3n dimensional non symplectic Nambu manifold $M^{3n}$ where a
nondegenerate closed 3-form takes
the place of the symplectic 2-form. A $C^\infty$ mapping $F$ :
$M^{3n} \to N^{3n}$ is a
canonical transformation if it leaves the 3-form invariant. In
particular, the Nambu equations
of motion realize a phase flow through two Hamiltonian functions
$h_1$ and $h_2$ with the time
evolution of the system being a unfolding of successive 3-form
preserving canonical
transformations. In such a Nambu system, a Poisson bracket of 2-forms closing
on an SO(3) algebra is the counterpart of the Poisson bracket of 1-forms or
vector fields of usual Hamilton dynamics. On the other hand
the triple brackets for functions { F, G, H} has a non-associative
structure as observed by \cite{nambu} and \cite{okubo}. There is a
Nambu-Darboux theorem whereby at every  point P in
$M^{3n}$ there is a coordinate chart (U,
$\phi\ )$ in which $\omega^3$\  has the Darboux form $\omega^{(3)} =
\sum_{i= 0} d x_{3i+1}
\wedge dx_{3i+ 2}
\wedge dx_{3i+3} $ (i=0,1,...n-1).

    It is specially noteworthy that the normalization of
the Nambu wave function or Born rule
$\sum _a {\bar{\Psi^{a}}} \Psi^{a} =const$ does not have to be imposed.
It is in fact
automatically conserved and given by the value of $h_2$ ! Thus the
usual normalization condition
for wave functions follows from the dynamical set-up and is strongly
indicative (after
appropriate normalization) of a
probabilistic or stochastic interpretation for the generalized Nambu
quantum mechanics.
On the other hand, the fact that the superposition principle may fail
suggests that the concept of measurement and the standard statistical
interpretation  has to be
reevaluated. The  condition
\eq
C = \sum_a [ (m_1^a)^2 + (m_2^a)^2 + (m_3^a)^2],
\en
where $C$ is a constant, defines the space of states of the Nambu
quantum mechanics together
with the requirement that the points in this space of states are
identified under the action of
$SO(3)$. In other words, $\Psi$ and
$U \Psi U^{\dagger}$, where $U$ is an $SO(3)$ matrix,
are identified as physical states. Thus there is a non-abelian
$SO(3)$ phase, which generalizes
the usual $U(1)$ phase of quantum mechanics. (It is a fascinating possibility
that the usual $U(1)$ phase could emerge dynamically from the
non-abelian $SO(3)$ phase.)  More
precisely, the normalization condition as in the U(1) phase case of
quantum mechanics, gives
rise to the structure of a SO(3) principal fibre bundle of spheres
over spheres, namely
$S^{3n+2} \to S^{3n-1}$ . We call these Nambu bundles and
readily check that these
bundles sit between the complex Hopf bundles
$S^{2n+1}\to CP^{n}$
of quantum mechanics
and the quaternionic Hopf bundles $S^{4n+3} \to HP^{n}$ of
quaternionic quantum mechanics, $HP^{n}$ being the n-dimensional (4n real)
quaternionic projective space. Namely that, for a given n, the complex line
bundles are embedded as they should within the Nambu bundles, which
are themselves
embedded in turn in the quaternionic line bundles. In this sense Nambu quantum
mechanics as formulated here is intermediate between complex and
quaternionic quantum
mechanics. The three quantum phase spaces are nested within one
another as $CP(n) \subset
    S^{3n-1} \subset HP(n)$. The Nambu manifold, a (3n-1)-sphere is
generally neither
symplectic nor complex. It is endowed with an invariant closed
nondegenerate 3-form
which descends from a closed 4-form, characteristic of $HP^n$ and
giving the latter a
quaternionic Kahler structure. The metric of
the "round" sphere $S^{3n-1}$ is the extension of the Fubini-Study
metric of the
complex quantum mechanics.
\eq
ds_{12}^2 = (1 - |<\Psi_1|\Psi_2>|^2).
\en
Finally, given the basic equations of the Nambu quantum mechanics, we see that
there exists a geometric phase generalizing the usual Berry phase.
The generalized geometric phase, which should be independent of the equation
of motion of first order in time, is given in terms of the Nambu wave
function,
\eq
\int \bar{\Psi} d\Psi.
\en

Clearly, many fundamental questions have yet to be answered about the above
Nambu quantum mechanics, the foremost one being its physical
interpretation. Thus the
bundle structure endowed with a closed nondegenerate 3-form suggests
the formulation
of a) some generalized Poisson structure and b) hypercomplex
structure for the Nambu
manifold, something intermediate between complex and quaternionic
structure. The existence of some kind of hypercomplex structure may yet
provide a nonlinear superposition principle, if any exist. This hope is
not out of reach as the classical Euler top is completely integrable with a
very apparent Lax structure. The fact that a generalized Born rule is built
in through the second Hamiltonian supports our view that
geometrizing Nambu mechanics is a natural step since it is $already$
quantized, that we are in the presence of a stochastic dynamical system
where the phase space
$S^{3n-1}$ is also a probability space. A n-sphere probability space
was considered
in \cite{Gibbons} in a spinorial generalization on quantum
mechanics.
Moreover, being the prototype Blaschke manifold
\cite{blaschke}, the round hyperspheres $S^n$ share with
the projective (Blaschke) spaces \cite{gursey}($RP^n$, $CP^n$, $HP^n$, $CaP^2$)
the defining property that their geodesics exhibit the most regular 
behavior. It is
known \cite{anandan, cirelli} that this regular
geodesic structure is at the basis of the generalized energy-time 
uncertainty relation.

One of the fundamental questions to be answered concerns
the nature and properties of the ``observables'' of Nambu quantum mechanics.
We know that they must be a rather restricted class of functions over
the Nambu manifold since in the symmetric top limit where standard quantum
mechanics is recovered, they must coincide with the Kahler functions.
The latter are in one to one correspondence with
observable hermitian
operators. Obviously the technology of the operatorial formalism for our scheme
has to be developed further and applied to simple physical systems \cite{vt}.
Similarly, the physical meaning of the
generalized uncertainty relations should be understood.
Another question is whether there exists a
Weyl-Wigner-Moyal-like deformation
quantization \cite{moyal}
formulation of geometric Nambu quantum  mechanics? We
hope to address some of these issues in future work.

Finally, we believe that the correct arena for the application of
our formalism is to be found in string theory.
In particular, in references \cite{stu} and
\cite{mthnambu} it has been argued that the structure of the Nambu bracket
appears quite naturally in the problem of a covariant formulation of
Matrix theory \cite{matrix} based on the analogy with the eleven-dimensional
membrane \cite{membrane}. One of the obstacles in this approach was
rooted in the quantization problem of the Nambu bracket. (Similarly the
quantization of the topological open membrane is closely related
to the problem of the quantization of the triple Nambu bracket \cite{boris}.)
As pointed out
in this article, our geometric formulation of the Nambu bracket allows
for an identification of the classical Nambu bracket with the quantum
Nambu bracket provided one works on the
configuration space of the Nambu quantum mechanics.
Thus, we believe a stage is set for an application of the Nambu bracket to
Matrix theory.

{\bf Acknowledgments:}
We thank V.~Balasubramanian, L. N. Chang, M. di Ventri, M. Gunaydin, T. 
Hubsch, F. Larsen, T. Mizutani,  N. Okamura,
B. Schmittmann, A. Sen, J. Slawny, U. Tauber, T. Takeuchi, and Royce Zia
for useful discussions. D.M. thanks to H. Awata, M. Li and T. Yoneya as well
as B. Pioline for many
discussions on Nambu dynamics.

\end{document}